\documentclass[aps,prl,showpacs,amsmath,amssymb]{revtex4}

\usepackage{graphicx}
\usepackage[latin2]{inputenc}

\usepackage{amsfonts}

\begin{document}

\title{Entanglement swapping in a Franson  interferometer setup}

\author{Ugo Messina and G. Massimo Palma}
\affiliation{ NEST - CNR (INFM) \& Dipartimento  di Scienze Fisiche ed Astronomiche, Universit\`a degli Studi di Palermo, via Archirafi 36,
I-90123 Palermo, Italy}

\date{10 February 2007}

\begin{abstract}
We propose a simple scheme to swap the non  local correlations, characteristic of a Franson interferometric setup, between pairs of frequency entangled photons emitted by distinct non linear crystals in a parametric down conversion process. Our scheme consists of two distinct sources of frequency entangled photons. One photon of each pair is sent to a separate Mach - Zender interferometer while the other photons of the pairs are mixed by a beam splitter and then detected in a Ou - Mandel interferometer. For suitably postselected joint measuremetns, the photons sent at the Mach -Zender show a coincidence photocount statistics which depends non locally on the settings of the two interferometers.
\end{abstract}

\maketitle

\section{introduction}
The presence of non local correlations in the joint photocounts probabilities of frequency entangled photons in two spatially separated Mach - Zender interferometers was detected by Franson more than a decade ago\cite{Franson1, Franson2}.  In his experimental setup, widely known now as Franson interferometer, two frequency entangled photons emitted by a two photon cascade or a type I down conversion process were sent to the input of two spatially separated Mach - Zender (M-Z) interferometers. The joint photocount statistics showed non local correlations with strong similarities to the ones shown in test of Bell inequalities. Such example of higher order interference has been since after widely analyzed both theoretically and experimentally \cite{Campos, Kwiat,Mandel} and has been proposed as a scheme for the implementation of quantum cryptographic protocols \cite{Rarity,Gisin}.

Given two pairs of entangled systems, $a,b$ and $a',b'$ it is possible to generate entanglement between systems  $a,a'$ by a suitable joint measurement on systems $b,b'$. Such scheme is known as entanglement swapping and it has been first proposed in \cite{swap}.
In the present paper we suggest an experimental scheme for the implementation of entanglement swapping between two Franson interferometers.
Our scheme consists of two type I sources of pairs of downconverted frequency entangled photons. One photon for each pair is sent to a M-Z interferometer while the two remaining photons are mixed at a beam splitter and then detected as in a typical Ou-Mandel interferometer \cite{Ou}. We show that for suitably postelected joint measurements of the photons leaving the beam splitter, the joint measurements at the two M-Z show  non local correlations similar to the ones characteristic  of the Franson Interferometer.
An experiment in a similar spirit has been carried out with time bin entangled photons \cite{Timebin}.
We will show that our scheme, closer to the original Franson setup, although requires brighter sources of entangled photons, requires less synchronization.

In the next section  we will review briefly, for the sake of completeness, the properties of our frequency entangled two photon state and of the Franson interferometer while in section \ref{ugo} we will illustrate our proposal.

\section{The Franson interferometer}

The output state of a the signal and idle modes of a type I parametric down conversion process can approximately written as 
\begin{equation}
\label{stato emesso}
\vert\psi\rangle \approx \int d\omega d\omega'\,
f(\omega,\omega')\vert\omega\rangle_a\vert \omega'\rangle_b
\end{equation}
where  $a$ and $b$ label two particular wavevector direction. If the nonlinear crystal is pumped at a frequency $2\Omega$ the probability amplitude function $f(\omega, \omega')$ shows pairwise entanglement between the modes around the frequency $\Omega$ and  takes the form
\begin{equation}
\label{effe}
f(\omega,\omega')\approx f(\omega)\delta(\omega+\omega'-2\Omega)
\end{equation}
where, to a good approximation \cite{Campos},
\begin{equation}\label{f gauss}
f(\omega) =  f_0 \exp\left\{-\frac{(\omega-\Omega)^2}{4\Delta\omega^2}\right\}
\end{equation}
The state (\ref{stato emesso}) can therefore be written as
\begin{eqnarray}
|\psi \rangle &=& \int d\omega\, f(\omega)\vert\omega\rangle_a\vert
2\Omega-\omega\rangle_b
\label{stato emesso in omega}\\
&=&\iint\, dt\,dt'\,\vert t\rangle_a\vert
t'\rangle_b e^{-2i\Omega t'}F(t-t')
\label{stato emesso in tt'}
\end{eqnarray}
where $F(t)=\int d\omega f(\omega ) e^{i\omega t}$ is the Fourier transform of $f(\omega)$. If we assume that the bandwidth $\Delta \omega$ of populated frequencies is large, the photon wavefunction becomes
\begin{equation}
\label{state in t}
|\psi \rangle \approx  f_0\int dt e^{-2i\Omega t}\vert
t\rangle_a\vert t\rangle_b
\end{equation}
The physical interpretation of Eqs.(\ref{stato emesso in tt'},\ref{state in t}) is straightforward: the pairs of correlated photons are emitted with a constant probability amplitude. Once a photon in one mode - say $a$ - is detected at time $t$, the probability amplitude to detect a photon  in mode $b$ collapses to a packet of time width $\tau \propto \Delta\omega^{-1}$. In the broadband limit $\tau \approx 0$, i.e. the photons are emitted in simultaneous pairs.

\begin{figure}
\begin{center}
\includegraphics[height=5.cm]{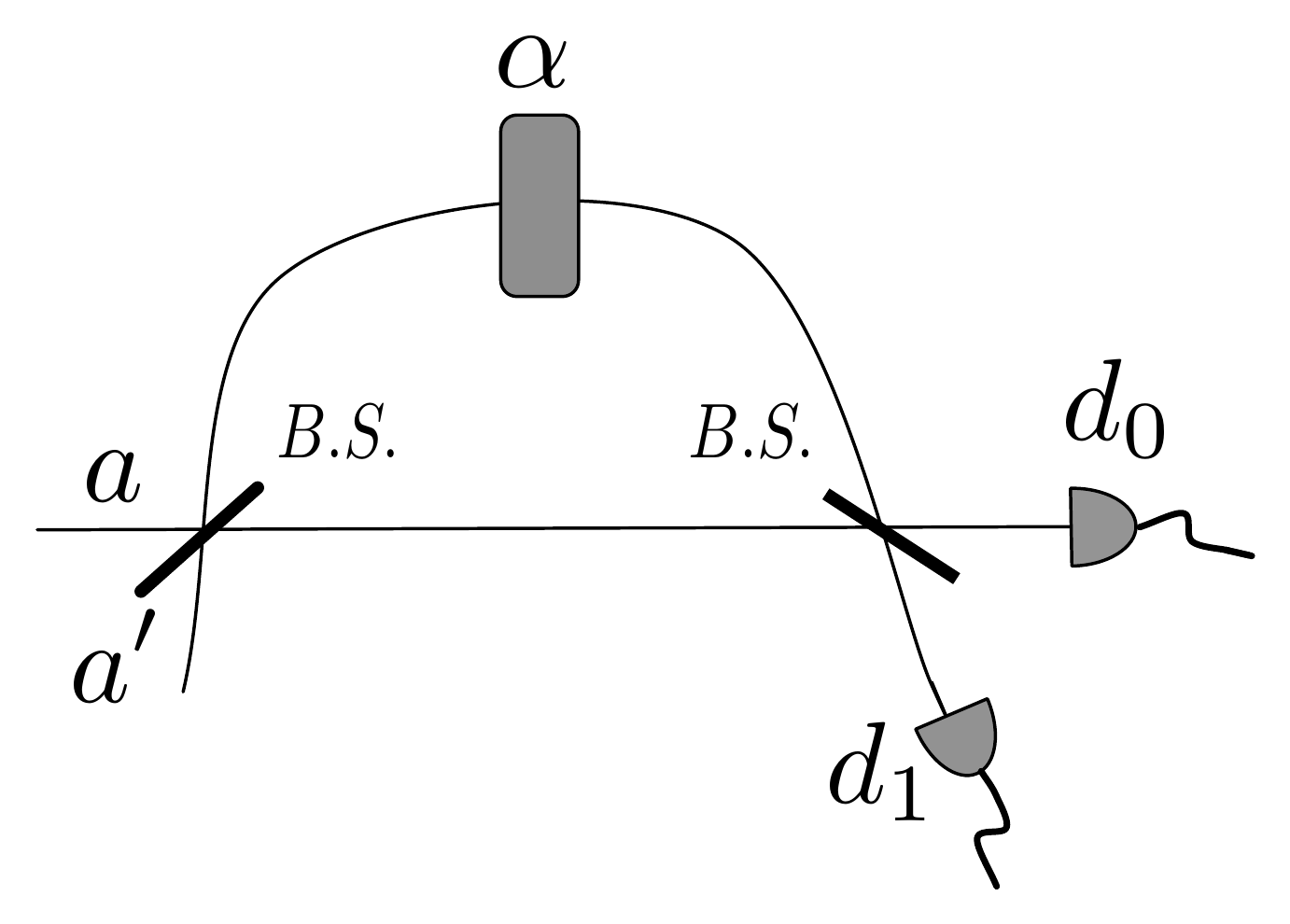}
\caption{Schematic reprresentation of a Mach-Zender interferometer with two input modes, $a$ and $a'$ and two output modes $d_0$ and $d_1$. The modes are mixed by symmetric $50\%/50\%$ beam splitters. The two arms of the interferometer have different optical length $S=ct_s$ and $L=ct_l$ respectivelly. An additional phase shift $\alpha$ can be experimentally introduced }
\label{MZ}
\end{center}
\end{figure}

\begin{figure}
\begin{center}
\includegraphics[width=9cm]{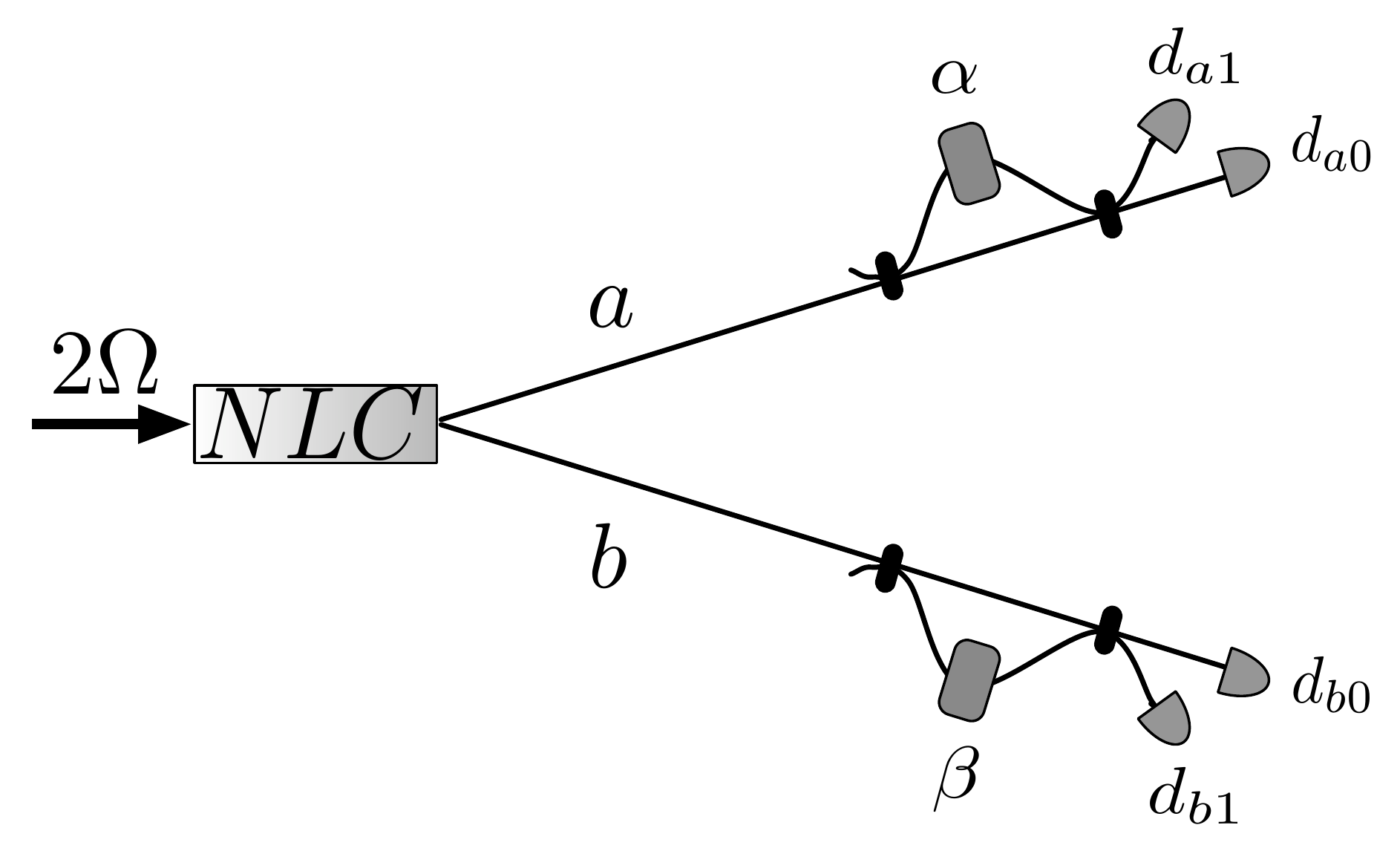}
\caption{ The photons leaving the non linear crystal, pumped by a strong laser at frequency $2\Omega$ are injected into two spatially separated M-Z interferometers }
\label{Franson interferometer}
\end{center}
\end{figure}
\begin{figure}
\begin{center}
\includegraphics[height=7.cm]{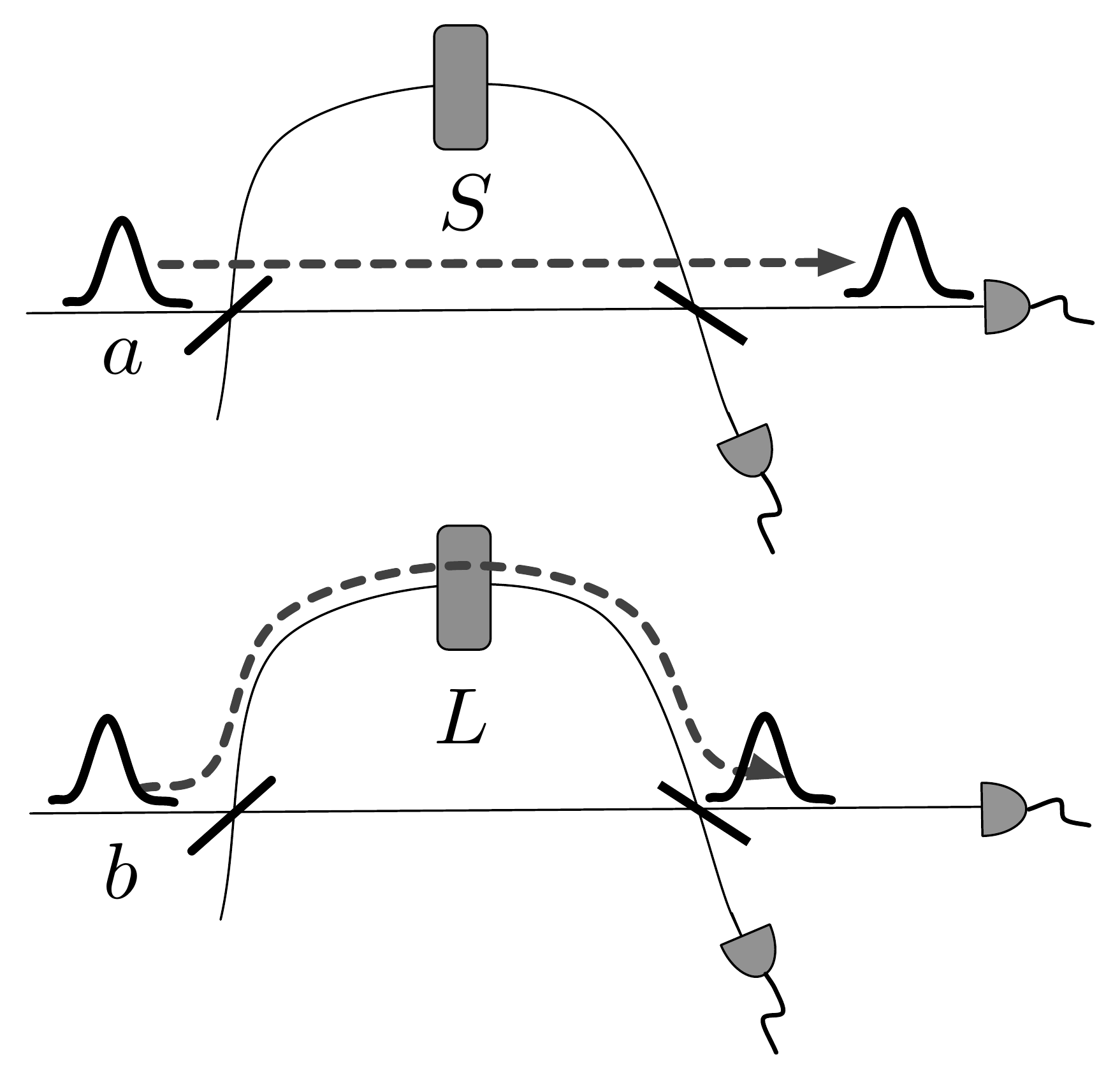}\hspace{0.0cm}
\includegraphics[height=7.cm]{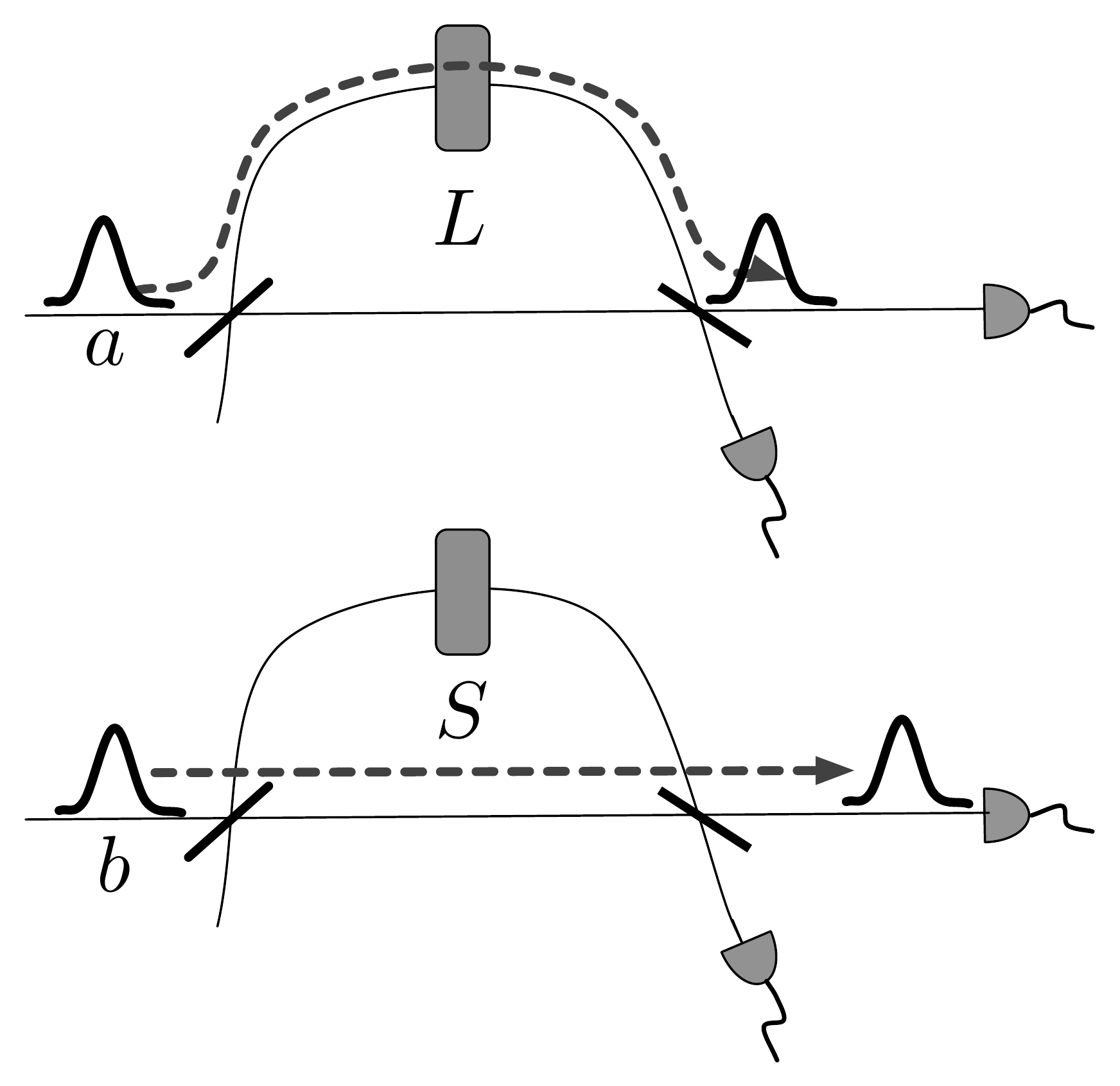}
\vspace{0.0cm}
\caption{When the pair of photons reach the photodetectors along the LS or SL paths one observes two distinguishable photodetection events separated by a time interval $\Delta t = t_l -t_s$ }
\end{center}
\label{ShortLong}
\end{figure}
\begin{figure}
\begin{center}
\includegraphics[height=7.cm]{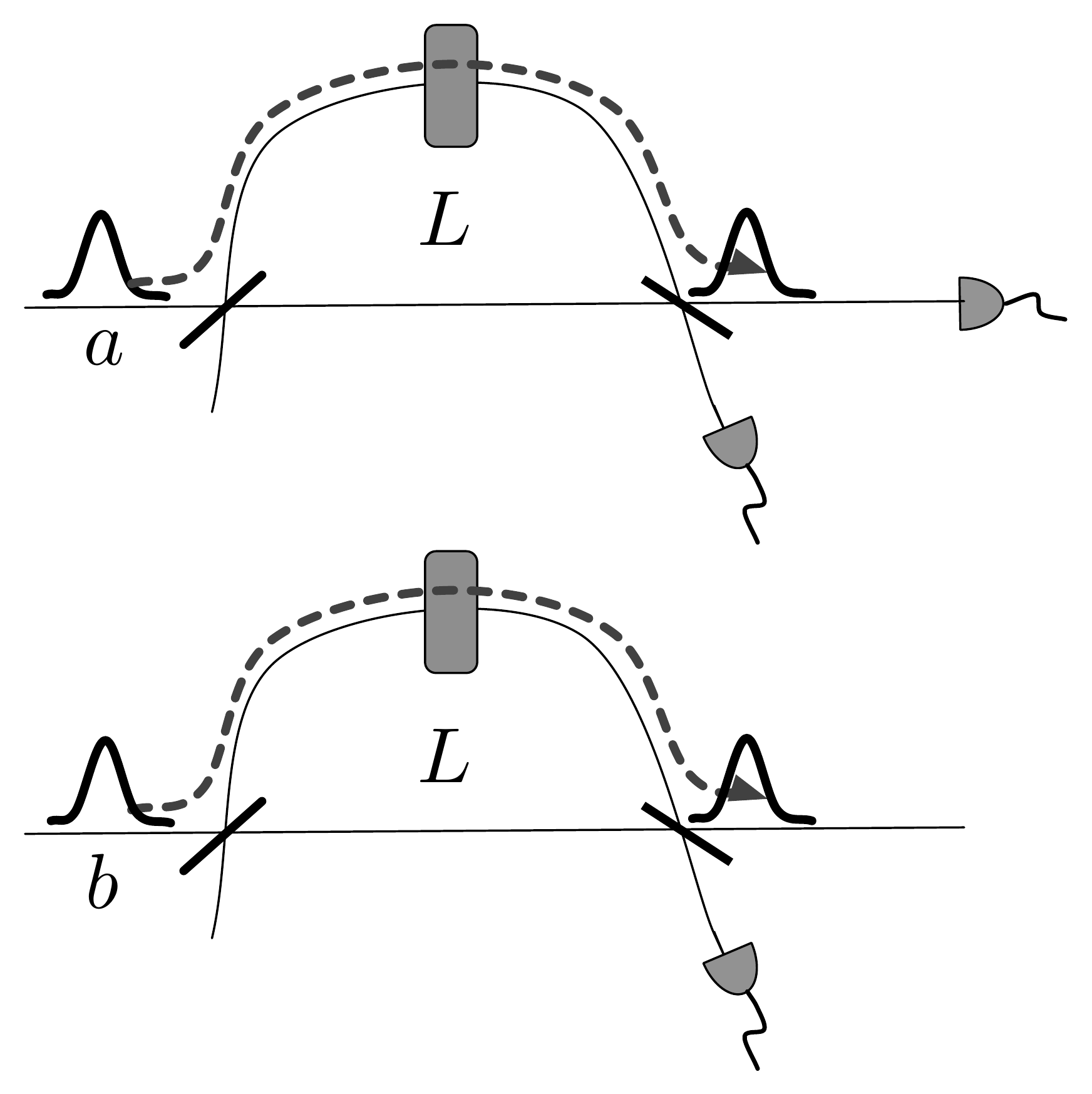}\hspace{0.0cm}
\includegraphics[height=7.cm]{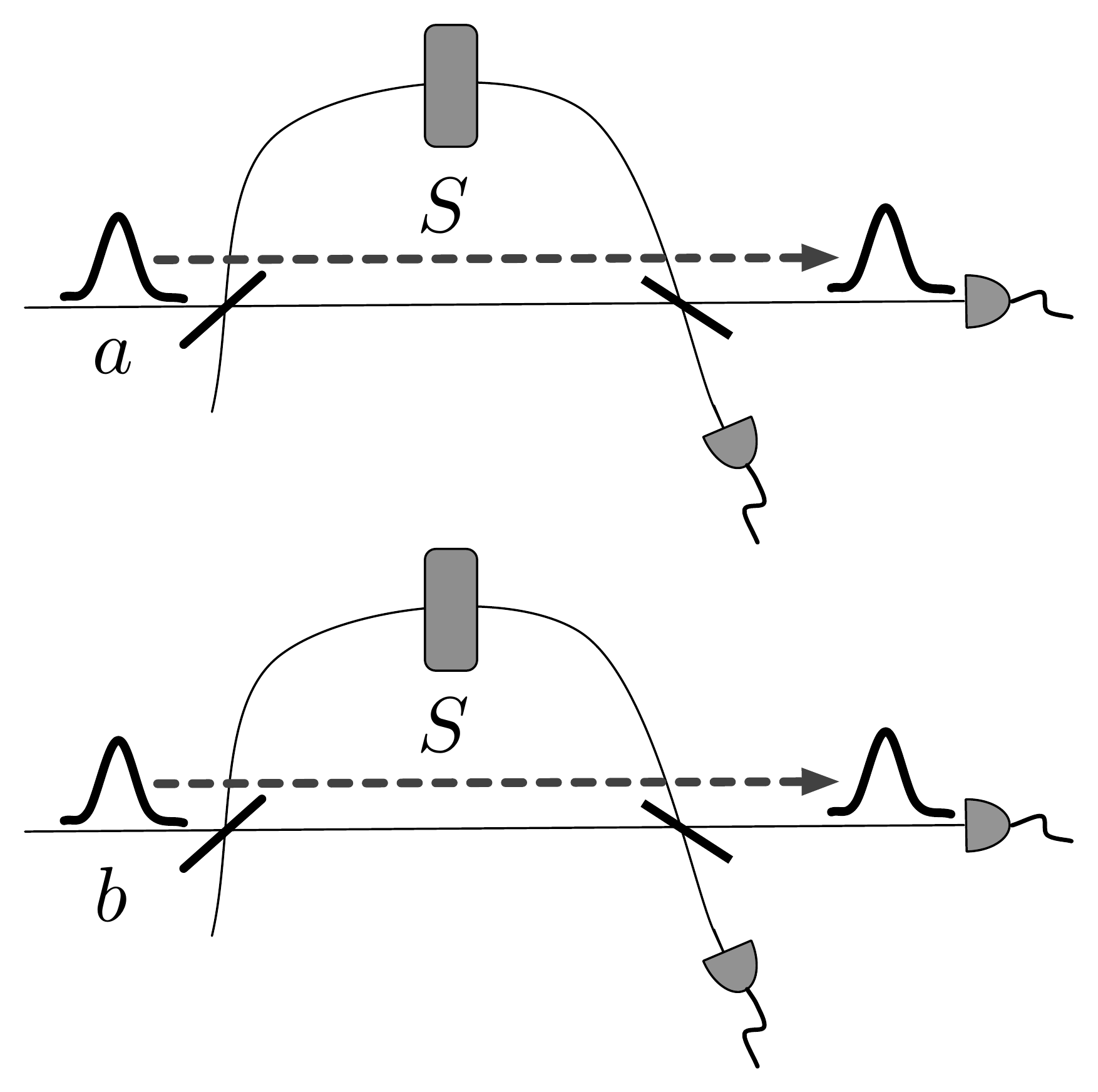}
\end{center}
\caption{In the case of simultaneous joint photodetections the photons may have reached the photodetectors either along the LL paths or the SS path. Such alternatives are indistinguishable and therefore can give origin to interference phenomena}
\label{LL}
\end{figure}

Assume now that the photons in mode $a$ and $b$ are injected in one of the input of two spatially separated  M-Z interferometers, as shown in Fig.(\ref{Franson interferometer}).  Writing (\ref{stato emesso in omega} )as

\begin{equation}
\label{ }
|\psi \rangle = \int d\omega f(\omega ) a^{\dagger} (\omega ) b^{\dagger}(2\Omega - \omega) |0\rangle_a |0\rangle_b
\end{equation}
and using the input-output relation at each of the Mach Zender interferometers, which can be straightforwardly deduced by inspecting Fig.(\ref{MZ})

\begin{align}
a^{\dagger}(\omega)&=\frac{1}{2}\left\{(e^{i\omega
t_s}-e^{i\omega t_l+i\alpha})d^{\dagger}_{a0}(\omega)+i(e^{i\omega
t_li+\alpha}+e^{i\omega
t_s})\hat{d}^{\dagger}_{a1}(\omega)\right\}\nonumber\\
&= \left\{c_{a0}(\omega)d^{\dagger}_{a0}(\omega)+c_{a1}(\omega)d^{\dagger}_{a1}(\omega)\right\}\\
b^{\dag}(\omega)&=\frac{1}{2}\left\{(e^{i\omega
t_s}-e^{i\omega t_l+i\beta})d^{\dagger}_{b0}(\omega)+i(e^{i\omega
t_l+i\beta}+e^{i\omega
t_s})\hat{d}^{\dagger}_{b1}(\omega)\right\}\nonumber\\
&= \left\{c_{b0}(\omega)d^{\dagger}_{a0}(\omega)+c_{b1}(\omega)d^{\dagger}_{b1}(\omega)\right\}
\end{align}
the wave functions (\ref{stato emesso in omega})  can be expressed in terms of the output modes of the two interferometers, labeled as $d_{a0}$, $d_{a1}$ and $d_{b0}$, $d_{b1}$ and takes the forms

\begin{equation}\label{4-stato out MZ}
\vert\psi\rangle=\int d\omega d\tilde{\omega}
f(\omega)\sum_{ij=0,1}c_{ai}(\omega)c_{bj}(\tilde{\omega})
\vert\omega\rangle_{ai}\vert \tilde{\omega}\rangle_{bj}\delta (\omega + \tilde{\omega} - 2\Omega)
\end{equation}
From the above equation, using the standard photodetection theory,  it is possible to obtain the joint probability distribution
$P_{ij}(t,t')$ that a photon is detected at detector $d_{ai}$ at time $t$ and a photon at the output $d_{bj}$ at time $t'$. It is straightforward to see that either the two detectors $d_{ai}d_{bj}$ register a simultaneous event or they register two events
separated by delay time $\Delta t = t_l - t_s$. The simultaneous detections show a non local dependence on the phase settings of
the two interferometers. These results can be easily understood by noting that the two photons are emitted simultaneously and are
localized packets of duration $\tau \approx 0$.  Each photon can reach the photodetectors via two possible paths, namely the long ($L$) and the short  ($S$) arm of the interferometer. To each path is attached a probability amplitude
with a phase factor which depend on the optical path and on the local phase. The photons can reach the  photodetectors $d_{ai}d_{bj}$ along the following paths: $L_a,S_b$,   $L_b,S_a$, Fig.( \ref{ShortLong} ) and  $L_a,L_b$,   $S_a,S_b$ Fig. (\ref{LL}). Since the wavepackets are well localized, i.e. since $\tau \ll  \Delta t$,  the $LS$ and $SL$ paths are distinguishable and are responsible of detection events separated by a time interval $\Delta t$.  As a consequence

\begin{equation}
P_{ij}(t, t\pm \Delta t) =\frac{1}{16}
\end{equation}

On the other hand the paths $LL$ and $SS$, which are responsible of simultaneous joint photodetections (see Fig. \ref{LL}),  are indistinguishable since, although it is known that the two photons enter
simultaneously the two interferometers, their exact emission time is unknown. The consequence of such indistinguishability is
interference between the probability amplitudes associated to such paths. A straightforward calculation shows that the probabilities of joint
simultaneous photocounts are

\begin{align}\label{joint1}
P_{00}(t,t)=P_{11}(tt)=\frac{1}{8}\left(1+\cos(2\Omega\Delta
t+\alpha+\beta)\right)
\\
P_{01}(t,t)=P_{10}(tt)=\frac{1}{8}\left(1-\cos
(2\Omega\Delta t+\alpha+\beta)\right)
\label{joint2}
\end{align}

Note how the simultaneous joint photocounts (\ref{joint1},\ref{joint2}) show a non local dependence of the local phase settings of the two spatially separated interferometers strongly reminiscent on the joint probabilities characteristic of Bell - inequality test experiments. As mentioned his has suggested the possible application of the Franson setup for quantum cryptographic applications \cite{Rarity, Gisin}.

\section{entanglement swapping scheme}
\label{ugo}

In this section we will show how such non local correlations can be swapped between  two distinct pairs of entangled photons which never
interacted. The key idea of entanglement swapping \cite{swap} is the following: given two separate maximally entangled pair of particles $a,b$ and $a' , b' $, if we perform a joint
Bell mesurement on particles $b,b'$, then particles $a,a'$ are projected in a maximally entangled state, although they never
interacted in the past. Following the above idea consider two independent non linear crystals emitting separate pairs of frequency entangled photons.  One may wonder if by means of a suitable joint measurement on pairs of
photons, each emitted by a separate source,  it is possible to reproduce the nonlocal Franson interference pattern with the remaining two
photons. We will show that this is indeed possible. 
\begin{figure}
\begin{center}
\includegraphics[width=9cm]{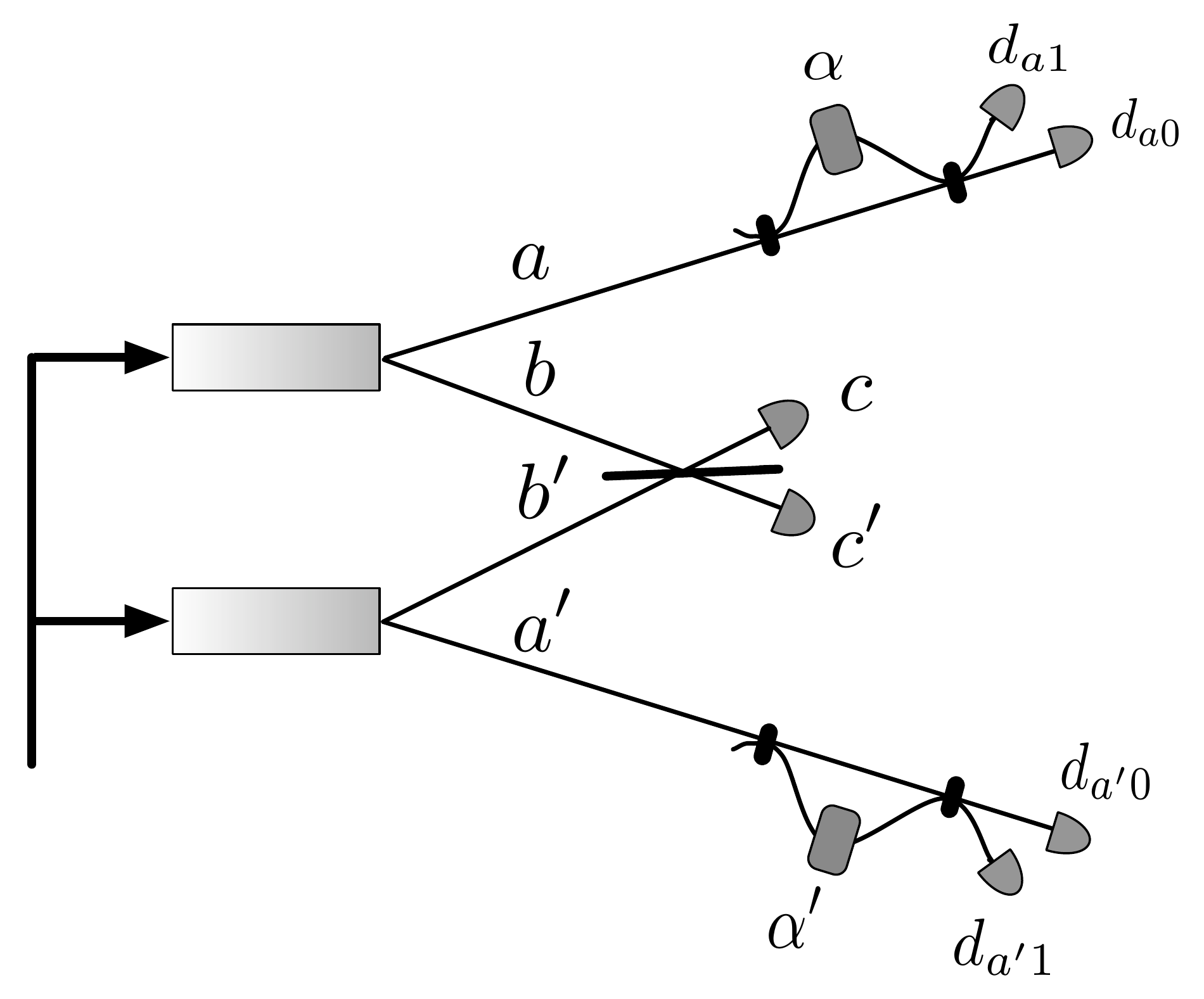}
\caption{Two separate non linear crystals are pumped by a strong laser at frequency $2\Omega$ and emit independent pairs $a,b$ and $a',b'$ of entangled photons. Photons $b$ and $b'$ are mixed at a symmetric beam splitter and the detected by detectors $c$ and $c'$}
\label{setup}
\end{center}
\end{figure}
In Fig.(\ref{setup}) is sketched the proposed setup. A strong pump field at frequency $2\Omega$ stimulates the spontaneous emission of pairs of parametric downconverted photons $a,b$ and $a',b'$ by two separate crystals. Photons in modes  $b$ and $b'$ are mixed at a beam splitter and detected - i.e. are sent in a Ou Mandel interferometer \cite{Ou}, while photons in modes $a$ and $a'$ are sent to two separate M - Z interferometers. 

\begin{figure}
\begin{center}
\includegraphics[width=9cm]{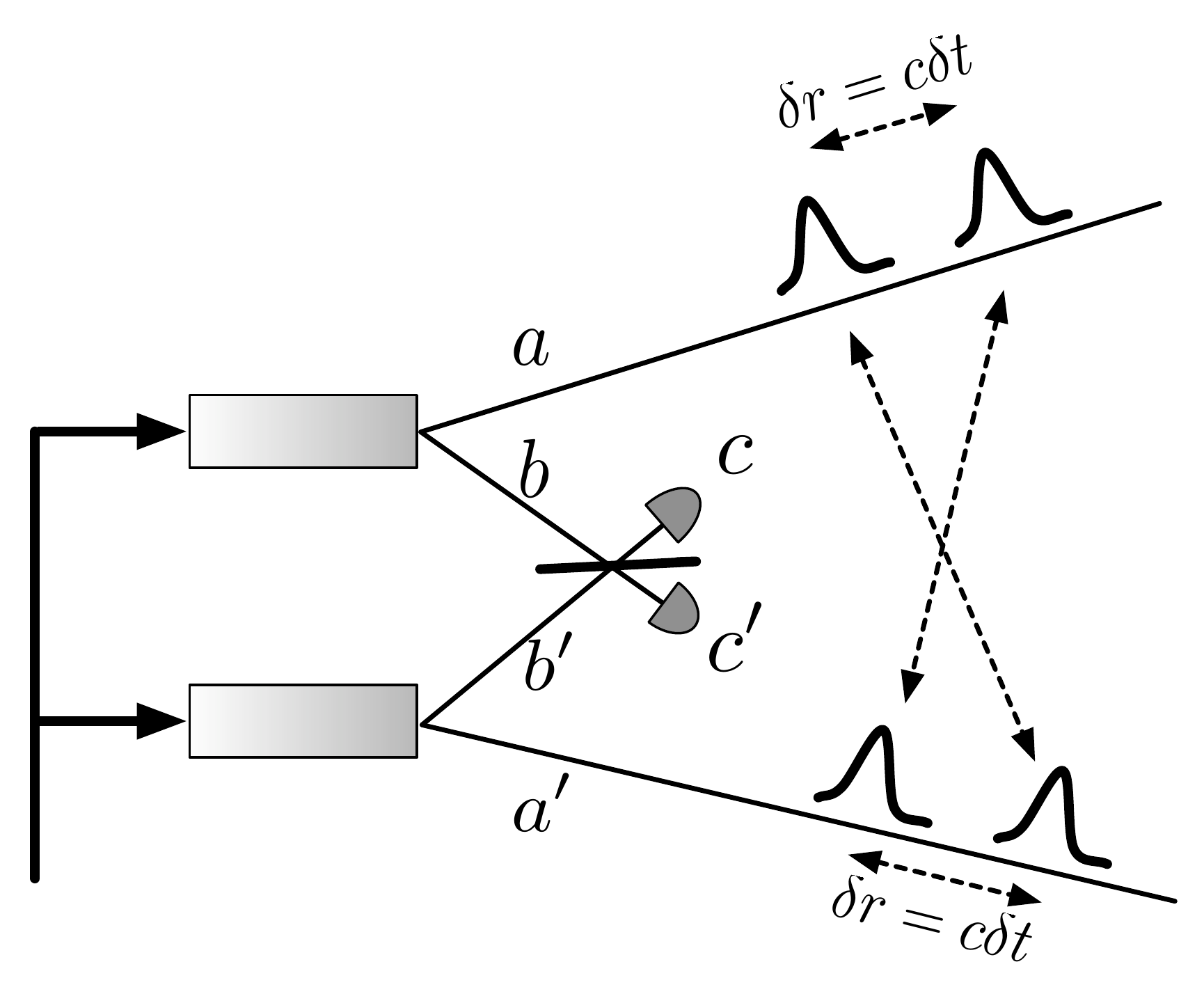}
\caption{ Two photons are detected in $c,c'$ at $t$ and $t+\delta t$. This collapses the wavefunction of modes $a,a'$ into a coherent superposition of two wavepackets separated by a time interval $\delta t$ }
\label{project}
\end{center}
\end{figure}

Let us first give a pictorial intuition of the effect of the measurement on photons $b,b'$ on the state of photons $a,a'$.  Suppose that two photons are detected in $c,c'$ at a time interval $\delta t$ and assume $ \delta t \gg \tau$ in order to neglect any bunching in the photocounts.  As we have discussed in the previous section  any measurement on one photon of an entangled pair localizes in time the other photon of the pair. In other words detecting photon $b$ at time $t$ gives information on the emission time of the entangled pair and therefore of photon $a$. However, since a photon detected in $c$ or in $c'$ could have come from $b$ or $b'$, it is not known if the collapse of the photon wavefunction took place in mode $a$ or in mode $a'$. When two photons are detected at time $t$ and $t+\delta t$ it is certain that both the state of mode $a$ and $a'$ have collapsed, but it is not possible to know in which order. Therefore the state of modes $a,a'$ is a coherent superposition of two wavepackets at a distance $\delta t$, as shown in Fig.(\ref{project}). In more mathematical terms, the wavefunction of the two pairs of photons is

\begin{equation}\label{stato iniziale 2S}
\begin{split}
\vert\psi\rangle &=\int d\omega\,f(\omega)\vert \omega \rangle_b\vert 2\Omega - \omega\rangle_a\otimes\int
d\omega'\,f(\omega')\vert\omega'\rangle_{b'}\vert 2\Omega - \omega'\rangle_{a'}\\&=
\iint d\omega d\omega'
f(\omega)f(\omega')\vert\omega,\omega'\rangle_{bb'}\vert 2\Omega - \omega,2\Omega - \omega'\rangle_{aa'}\\
&=\frac{1}{2}\iint d\omega
d\omega'
f(\omega)f(\omega')\Bigl(i\vert\omega,\omega'\rangle_{cc}+
\vert\omega',\omega\rangle_{cd}-\vert\omega,\omega'\rangle_{cd}+i\vert\omega,\omega'\rangle_{dd}\Bigr)
\vert 2\Omega - \omega,2\Omega - \omega'\rangle_{aa'}
\end{split}
\end{equation}
where in the last equation we have expressed modes $b,b'$ in terms of the photodetector modes $c,c'$. A straightforward calculation shows that when two photons are detected at time $t$ and $t+\delta t$ on the same photodector ($c,c$ or $c',c'$), the wavefunction of the photon pair in $a,a'$ collapses to

\begin{equation}\label{post-abs cc}
\vert\Psi^{+}\rangle =\frac{1}{\sqrt{2}}\Bigl(\vert t+\delta
t\rangle_a\vert r\rangle_{a'}+\vert t\rangle_a\vert t+\delta
t\rangle_{a'}\Bigr)
\end{equation}
while if the photons are detected in different photodetectors, $c,c'$ and $c',c$ the wavefunction of modes $a,a'$ collapses to 
\begin{equation}\label{post-abs cd}
\vert\Psi^{-}\rangle= \frac{1}{\sqrt{2}}\Bigl(\vert
t\rangle_a\vert t+\delta t\rangle_{a'}-\vert t+\delta
t\rangle_a\vert t\rangle_{a'}\Bigr)
\end{equation}
In both cases we have a coherent superposition of two wavepackets separated by a time interval $\delta t$. Note incidentally that if $\delta t \approx \tau$ we observe buncing. Indeed in this case  $\langle t | t+\delta t\rangle \neq  0$,  $ \vert\Psi^{+}\rangle \rightarrow |t\rangle_a|t\rangle_{a'}$ and $\vert\Psi^{-}\rangle \rightarrow 0$, i.e. we do not have an entangled  superposition of localized photon wavepackets.

\begin{figure}
\begin{center}
\includegraphics[height=5cm]{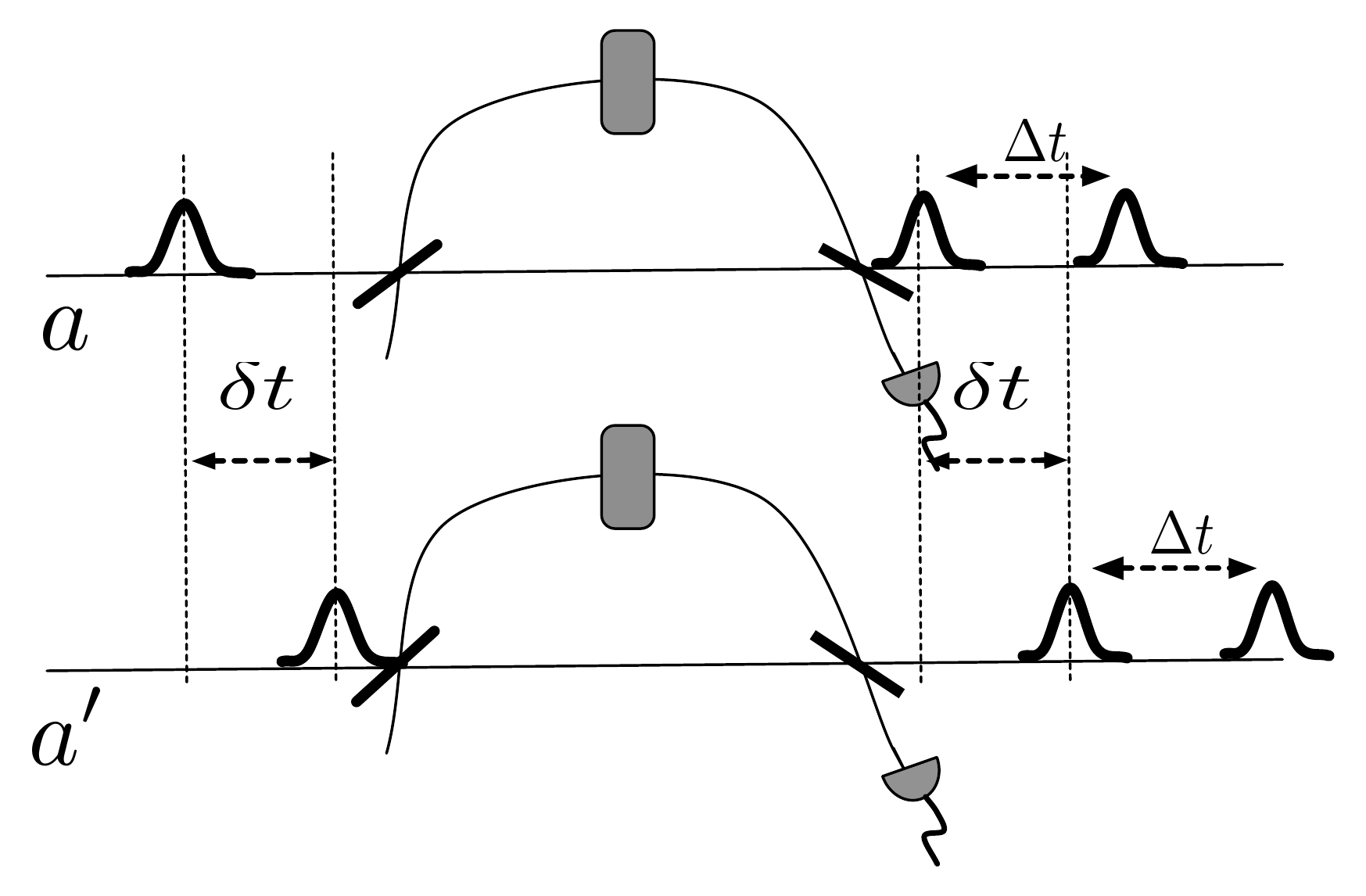}\hspace{1 cm}
\includegraphics[height=5.cm]{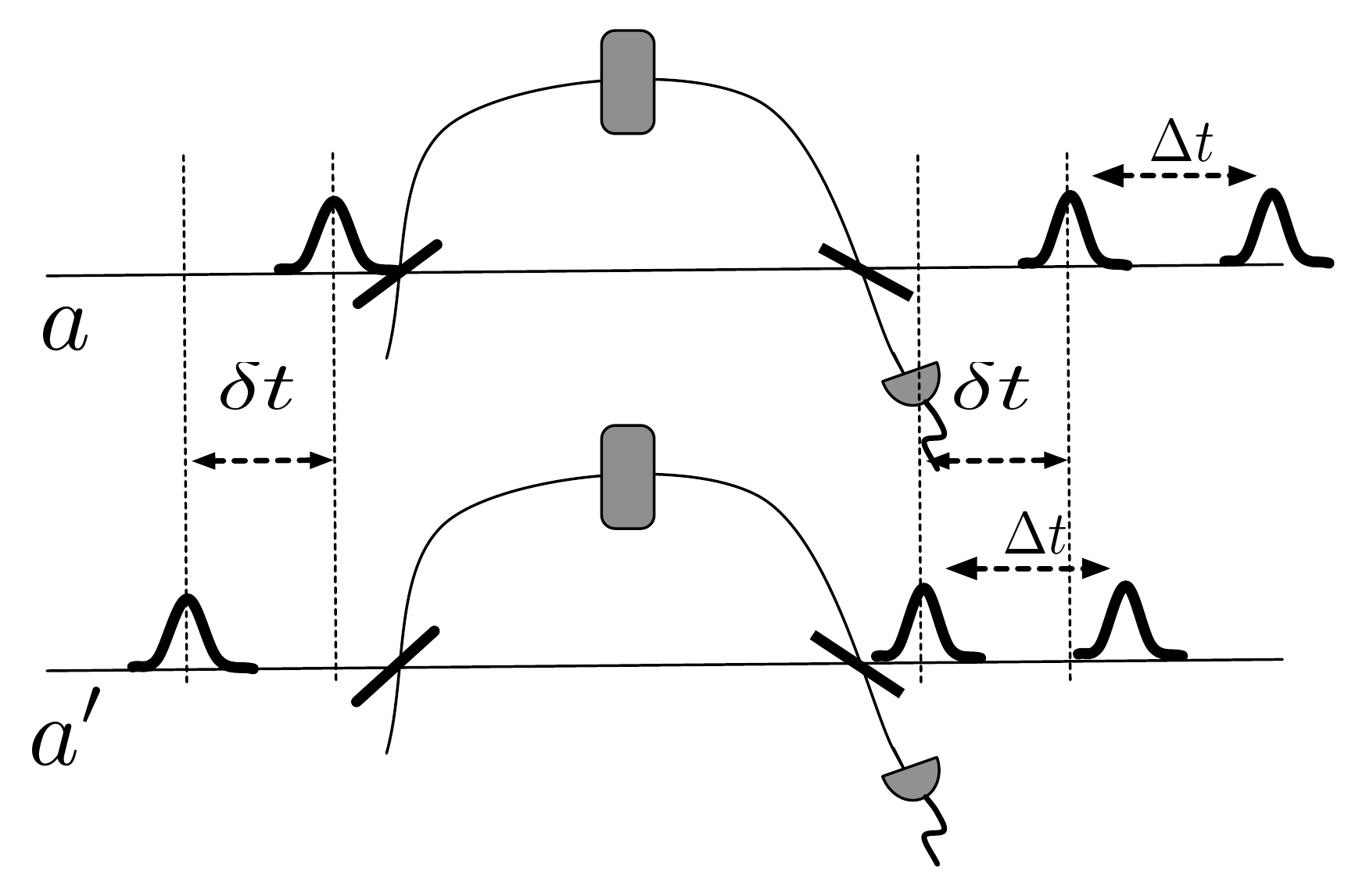}
\caption{A coherent superposition of the localized pulses in modes $a,a'$ shown at the left and at the right of the present figure enters the two MZ interferometers.  When the time delay between the input pulses $\delta t$ differs from the time delay $\Delta t = t_l - t_s$ the two sequences of output pulses are distinct an therefore distinguishable. For the sake of simplicity here we have drawn the sequence of pulses at detectors $d_{ao}$ and $d_{a'0}$. The same sequence of pulses reaches detectors $d_{a1}$ and $d_{a'1}$}
\label{input}
\end{center}
\end{figure}

\begin{figure}
\begin{center}
\includegraphics[height=5cm]{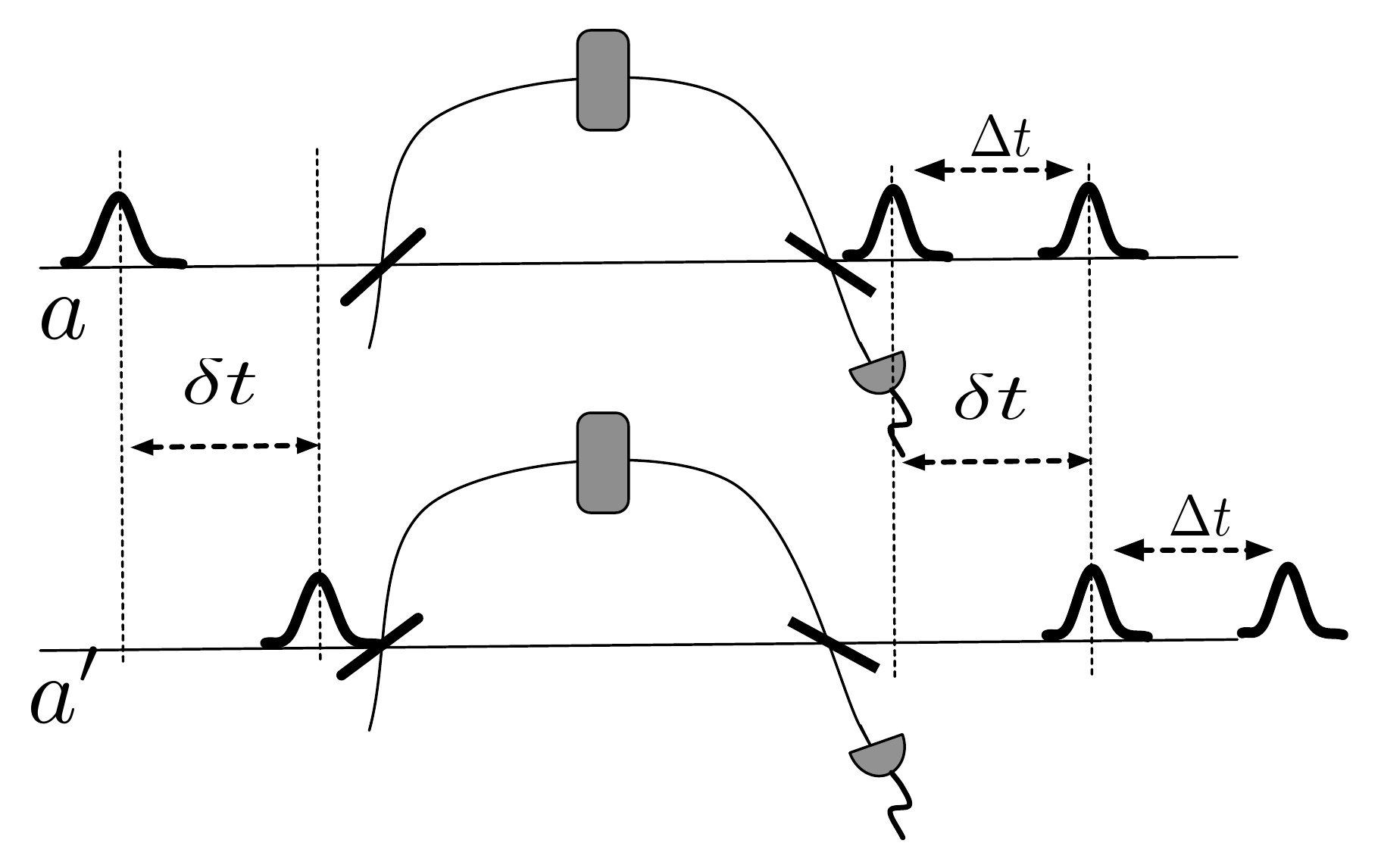}\hspace{1cm}
\includegraphics[height=5cm]{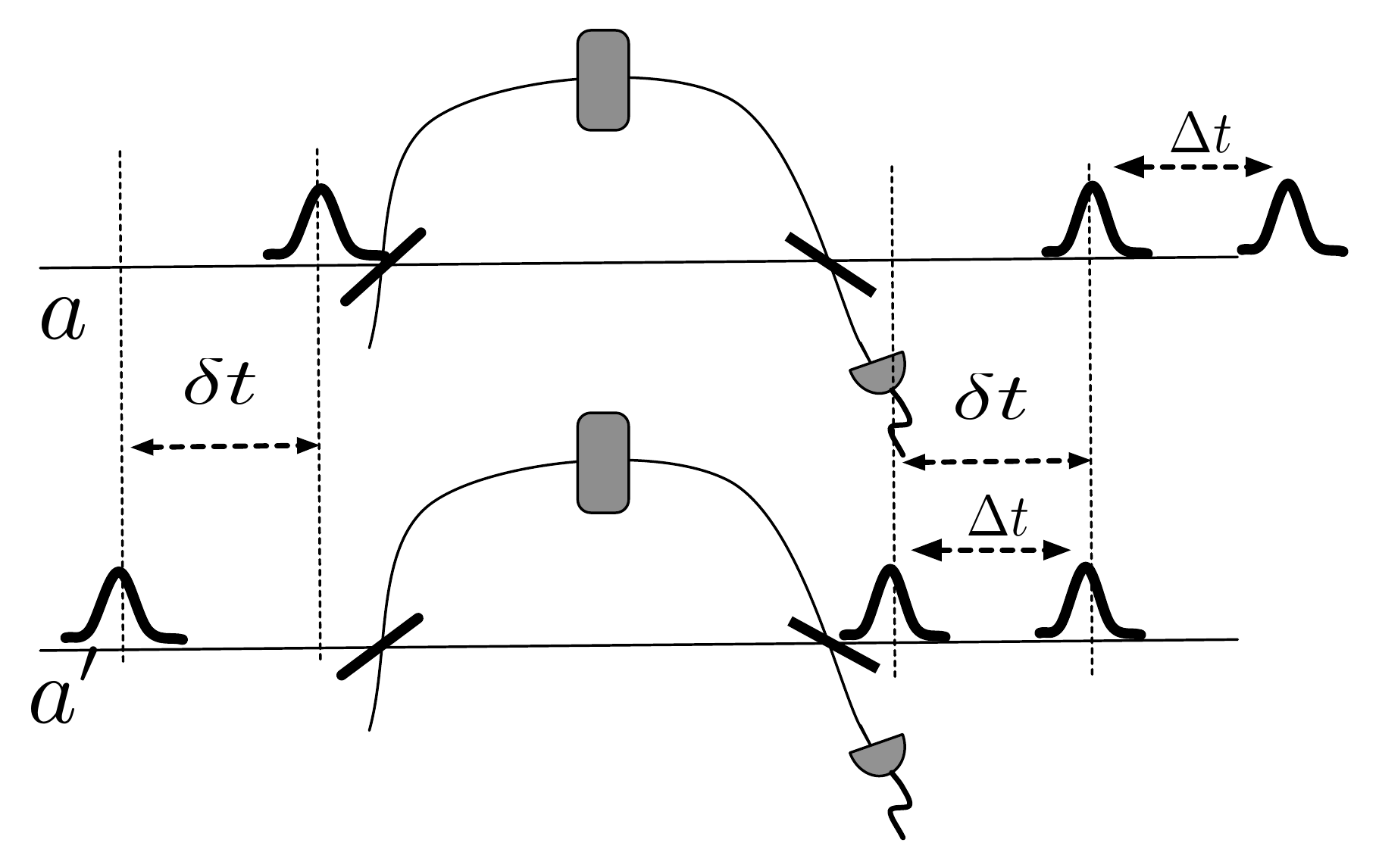}
\caption{A coherent superposition of the localized pulses in modes $a,a'$ shown at the left and at the right of the present figure enters the two MZ interferometers. When the time delay between the input pulses $\delta t$ equals the time delay $\Delta t = t_l - t_s$ the simultaneous  output pulse may originate either from the left or the right sequence of input pulse which are  indistinguishable. For the sake of simplicity here we have drawn the sequence of pulses at detectors $d_{ao}$ and $d_{a'0}$. The same sequence of pulses reaches detectors $d_{a1}$ and $d_{a'1}$}

\label{overlap}
\end{center}
\end{figure}

When a short single photon wavepacket enters a M-Z interferometer one observes a sequence of two pulses separated in time by an interval $\Delta t = t_l - t_s$ with equal probability at each of the two output detectors i.e. each pulse may reach the output either along the long or the short arm of the interferometer:
\begin{equation}\label{pacchetto out MZ in t}
\vert t\rangle\rightarrow\frac{1}{2}\{\vert
t+t_s\rangle_0-e^{i\alpha}\vert t+t_l\rangle_0+ie^{i\alpha}\vert
t+t_l\rangle_1+i\vert t+t_s\rangle_1\}
\end{equation}
In our case a coherent superstition of two wavepachets separated in time by $\delta t$ enters the two interferometers $a$ and $a'$. Since, as shown in Fig.(\ref{input}) the two sequences of output pulses which originate are distinguishable no nonlocal interference effect like the one described in the previous section can be observed. This is however is no no longer true when $\delta t = \Delta t$. In this case, as shown in Fig.(\ref{overlap}), some pulses may have reached the output via two indistinguishable paths and therefore one expects the appearance of a non local interference pattern similar to the one which characterizes the Franson interfereometer. In particular the event associated with pulse $a$ propagating along the short arm and pulse $a'$ propagating along the long arm is indistinguishable from the event associated with  pulse $a'$ propagating along the short arm and pulse $a$ propagating along the long arm. This leads to the following joint simultaneous photocount probabilities: 
\begin{align}
\label{P1}
P^{(+)}_{i=j}(t,t)&= \frac{1}{16}(1+\cos\Bigl((\alpha-\beta)\Bigr)\hspace{2cm}
&P^{(+)}_{i\neq j}(t,t)& = \frac{1}{16}\Bigl(1-\cos(\alpha-\beta)\Bigr)\\
P^{(-)}_{i=j}(t,t)&= \frac{1}{16}(1-\cos\Bigl(\alpha-\beta)\Bigr)\hspace{2cm}
&P^{(-)}_{i\neq j}(t,t)&=\frac{1}{16}\Bigl(1+\cos(\alpha-\beta)\Bigr)
\label{P2}
\end{align}
The above photocount probabilitie eq.(\ref{P1},\ref{P2}) show strong similarities and some differences with the photocout probabilities of the original Fraqnson experiment, Eq.(\ref{joint1})(\ref{joint2}). In both cases there is a non local dependence on the phase settings of the spatially separated MZ interferometers. In both cases the joint simultaneous photocounts are modulated by the phase difference between the  two interfering paths. In  (\ref{joint1},\ref{joint2}) such phase is $2\Omega\Delta t + \alpha +\beta$, i.e. the phase difference between the $LL$ and $SS$ while  in (\ref{P1})(\ref{P2}) the modulating phase is $\alpha - \beta$ i.e. the phase difference between the $LS$ and the $SL$ paths. The reason of such difference is that while in the original Franson setup the input photon pairs are delocalized in time, in our scheme they are localized in two wavepackets separated in time by $\delta t = \Delta t$. Furthermore the (\ref{P1})(\ref{P2}) depend on wether the input state is $|\Psi^{+}\rangle$ or $|\Psi^{-}\rangle$ i.e. on wether the photons at the Ou-Mandel interferometer are detected in the same detector on on different detectors.

\section*{acknowledgments}
G.M.P. acknowledges conversations with Dr. F. Bovino, Dr.
M. Genovese, Prof. A. Sergienko, Dr. H. Zbinden. This work has been supported by Italian MIUR under the funding PRIN 2006

\end{document}